\documentclass[aps,prl,amsmath,amssymb,floatfix,twocolumn]{revtex4}
\usepackage{graphicx}
\usepackage{subfigure}
\usepackage{color}
\usepackage{hyperref}

\begin{document}

\title{Chiral magnetic effect of Weyl fermions and its applications to cubic noncentrosymmetric metals}
\author{Pallab Goswami}
\affiliation{National High Magnetic Field Laboratory and Florida State University, Tallahassee, Florida 32310, USA}
\author{Sumanta Tewari}
\affiliation{Department of Physics and Astronomy, Clemson University, Clemson, SC 29634, USA}

\begin{abstract}
When the right and the left handed Weyl points are separated in energy, they give rise to a non-dissipative charge current along the direction of a uniform applied magnetic field, even in the absence of an external electric field.  This effect is known as the chiral magnetic effect and is a hallmark of the underlying chiral anomaly of the Weyl fermions.  According to the linearized continuum theory of Weyl fermions, the induced current is proportional to the magnetic field strength and the energy separation with a universal coefficient $e^2/h^2$. By considering a generic tight binding model for the cubic noncentrosymmetric metals, we show that such a system naturally supports a set of Weyl points, which are separated in energies. We also show the existence of the chiral magnetic effect for generic band parameters, and recover the universal result of the continuum Weyl fermions for a restricted parameter regime. Therefore, cubic noncentrosymmetric metals can serve as suitable platforms for realizing Weyl fermions and the exotic chiral elctrodynamic phenomena, which have promising technological applications.
\end{abstract}

\maketitle

{\em Introduction :} Recently it has been recognized that gapless fermionic systems can be endowed with non-trivial momentum space topology\cite{Volovik1}. The Weyl semi-metal is an example of such a gapless state, where the linearly dispersing left and  right handed Weyl fermions respectively appear as the unit charge monopole and anti-monopole of the Berry curvature\cite{Volovik1,Vishwanath}. For the topological non-triviality, it is essential that the Weyl points are separated in the Brillouin zone by a momentum $\delta \mathbf{K}$. The Berry flux passing through a plane perpendicular to $\delta \mathbf{K}$, which is located between the Weyl points is quantized to be unity, and determines the topological stability and the existence of protected chiral surface states. Therefore, the Weyl semi-metal is a unique fermionic quantum critical system with non-trivial topology. A number of systems have been recently shown \cite{Vishwanath,Xu,Burkov1,Burkov2,Zyuzin1,Zyuzin2,Cho,Meng,Gong,Sau} to support massless Weyl fermions (in either semi-metal or superconducting phases) as the bulk low energy excitations, and topologically protected gapless surface states in the form of open Fermi/Majorana arcs.For $\delta \mathbf{K}=0$, the monopole and the anti-monopole collapse onto each other, giving rise to the topologically trivial massless Dirac fermion, which describes the quantum critical point between the topological and the trivial band insulators \cite{Goswami1}. Due to the absence of Berry flux, there are no protected surface states for massless Dirac fermions.

Both semi-metallic phases (Weyl and massless Dirac) are examples of fermionic quantum critical systems with dynamic exponent $z=1$, and possess similar critical properties for various quantities such as the specific heat, the compressibility, the diamagnetic susceptibility, and the longitudinal conductivity, which are insensitive to the topology\cite{Goswami1,Hosur}. The topological distinction between the two systems can manifest through certain Berry curvature induced anomalous, non-dissipative transport properties, which can only be present in the Weyl semi-metal. For simplicity consider only one pair of Weyl fermions with momentum separation $\delta \mathbf{K}$, and energy separation $\Delta$. This system violates parity and time reversal symmetries (due to $\Delta$ and $\delta \mathbf{K}$ respectively), and the low energy Hamiltonian acquires the following form
\begin{equation}
\hat{H}_{W}=\begin{bmatrix}\frac{\Delta }{2}\sigma_0 + \hbar v\boldsymbol \sigma .\left(\mathbf{k}+\frac{\delta\mathbf{K}}{2}\right) & 0  \\ 0  & -\frac{\Delta }{2}\sigma_0- \hbar v\boldsymbol \sigma .\left(\mathbf{k}-\frac{\delta\mathbf{K}}{2}\right) \end{bmatrix},
\label{eq:1}
\end{equation}
where $v$ is the Fermi velocity, and the Pauli matrices $\boldsymbol \sigma$ and the $2 \times 2$ identity matrix $\sigma_0$ operate on the spin or the pseudo-spin indices of the two component Weyl spinors. The energy and the momentum separations respectively act as the temporal and the spatial components of a constant axial gauge potential. When this system is coupled to the electromagnetic gauge field, quantum field theory based calculations show the existence of two types of anomalous transport properties, which are tied to the axial anomaly of the Weyl fermions\cite{Zyuzin3,Grushin,Qi,Goswami2}. The momentum space separation leads to the anomalous Hall effect (AHE) in the absence of an external magnetic field, and the Hall current
\begin{equation}
\mathbf{j}_{H}=\frac{e^2}{2\pi h} \; \delta \mathbf{K} \times \mathbf{E},
\label{eq:2}
\end{equation}
where $\mathbf{E}$ is the external uniform electric field. On the other hand the energy separation leads to a charge current
\begin{equation}
\; \mathbf{j}_{ch}=\frac{e^2}{h^2} \; \Delta \; \mathbf{B}
\label{eq:3}
\end{equation}
along the direction of an applied uniform magnetic field $\mathbf{B}$, even in the absence of an applied electric field, which is known as the chiral magnetic effect (CME) \cite{Kharzeev,Landsteiner}. Both currents can be large and have potential technological applications.

In recent work based on a lattice model with one pair of Weyl fermions, it has been claimed that the CME is absent in real materials due to the lattice effects, even though the AHE formula of Eq.~(\ref{eq:2}) remains valid \cite{Franz}. The absence of the CME has also been claimed in a few other papers, which work with the continuum model of Weyl fermions, augmented by an ad hoc ultra-violet cutoff \cite{Basar,Landsteiner2}. These conclusions have been challenged in Ref.~\onlinecite{Burkov4}, where it has been shown that the answer for the CME in a linear response theory crucially depends on the order in which the external frequency and the wave-vector are sent to zero. In this Letter, we first show that both the AHE and the CME are the consequences of the electromagnetic gauge invariance in the presence of an axial gauge potential, and for a pair of Weyl fermions, separated in energy and momentum, both effects should exist. We demonstrate the finiteness of the CME for a lattice model of Weyl fermions, and also recover the universal result of Eq.~(\ref{eq:3}) for a restricted region of parameter space, when the lattice model can be mapped onto the linearized continuum theory. In particular, we adopt a tight binding model appropriate for describing cubic noncentrosymmetric metals such as LiPd$_3$B, LiPt$_3$B, and show that the spin-orbit coupling naturally gives rise to the Weyl points separated in energy space. In addition to establishing a nonzero CME for a lattice model, our results also propose the cubic noncentrosymmetric metals as suitable experimental platforms for realizing the Weyl semi-metal phase, as well as measuring the exotic CME.

{\em Axial anomaly and gauge invariance :} Before dealing with a lattice model, we first describe how the AHE and the CME are related to the axial anomaly of Weyl fermions. When the Weyl fermions are simultaneously coupled to the electrodynamic and the axial gauge fields respectively denoted by $\mathcal{V}_\mu$, $\mathcal{A}_\mu$, according to
\begin{eqnarray}
\mathcal{S}[\bar{\Psi},\Psi,\mathcal{V}_\mu, \mathcal{A}_\mu]= \int d^4x \bar{\Psi}i\gamma_\mu(\partial_\mu+ie\mathcal{V}_\mu+i\mathcal{A}_\mu \gamma_5)\Psi,
\end{eqnarray}
both the vector current $j^V_{\mu}=\bar{\Psi} \gamma_{\mu} \Psi$ and the axial current $j^A_\mu=\bar{\Psi}\gamma_{\mu}\gamma_5\Psi$ become non-conserved due to the axial anomaly \cite{Fujikawa,Qi,Goswami3}. We have defined the fourth coordinate $x_0=vt$. The violation of the conservation laws are described by
\begin{eqnarray}
&&\partial_\mu j^A_{\mu}= \frac{\epsilon^{\alpha \beta \rho \lambda}}{16 \pi^2}\left(e^2 F^V_{\alpha \beta}F^V_{\rho \lambda}+ F^A_{\alpha \beta}F^A_{\rho \lambda}\right),\\
&&\partial_\mu j^V_{\mu}=\frac{\epsilon^{\alpha \beta \rho \lambda}}{8 \pi^2}\; e \; F^V_{\alpha \beta}F^A_{\rho \lambda},
\end{eqnarray}
where $F^V_{\alpha \beta}$, $F^A_{\alpha \beta}$ respectively denote the field strength tensors of the electrodynamic and the axial gauge fields. However if we redefine the vector current as
\begin{equation}
j^V_{\mu} \to j^V_\mu-j^\prime_\mu=j^V_\mu- \frac{\epsilon^{\mu \nu \rho \lambda}}{4 \pi^2}\; e \; \mathcal{A}_\nu F^V_{\rho \lambda}
\end{equation}
the conservation law for the vector current (or the electromagnetic gauge invariance) can be restored. However, notice that the axial or the valley current always remains non-conserved for preserving the electromagnetic gauge invariance. The effects of anomaly on the axial current have been considered in Refs.~\onlinecite{Ninomiya,Aji}. For our model of Weyl fermions in Eq.~(\ref{eq:1}), the constant axial gauge potential $\mathcal{A}_\mu=(\Delta/(2 v), \delta \mathbf{K}/2)$, and $j^\prime_\mu$ leads to the results of Eq.~(\ref{eq:2}) and Eq.~(\ref{eq:3}), after restoring all the dimensionful units. Therefore, for a pair of Weyl fermions with $\Delta \neq 0 $, $\delta \mathbf{K} \neq 0$, the  observation of one effect and not the other constitutes a severe violation of the electromagnetic gauge invariance. This is one of the motivation behind our detailed analysis of the CME for a lattice model of Weyl fermions.

{\em Tight-binding model :} We consider the following tight binding Hamiltonian,
\begin{eqnarray}
&&H=\sum_{\mathbf{k}} \psi^\dagger_{\mathbf{k}}\left[N_{0,\mathbf{k}} \; \sigma_0 + \mathbf{N}_{\mathbf{k}} \cdot \boldsymbol \sigma \right]\psi_{\mathbf{k}}, \label{tightbinding} \\
&& N_{0,\mathbf{k}}= 2t_1\sum_{i}\cos k_i + 4t_2 \sum_{i<j} \cos k_i \cos k_j \nonumber \\
&& + 8t_3\prod_{i}\cos k_i+ 2t_4 \sum_{i}\cos 2k_i,\\
&& N_{j,\mathbf{k}}=t_{SO}\; \sin k_j,
\end{eqnarray}
which is applicable for describing the low energy band structure of various cubic noncentrosymmetric metals \cite{Takimoto}. In the above Hamiltonian, the $t_{SO}$ is the spin-orbit coupling strength and $t_1$, $t_2$, $t_3$, $t_4$ respectively denote the first, the second, the third and the fourth neighbor spin-independent hopping strengths. The Pauli matrices $\sigma_i$, and the $2 \times 2$ identity matrix $\sigma_0$ operate on the spin indices of the two component spinor $\psi^T_{\mathbf{k}}=(c_{\mathbf{k},\uparrow}, c_{\mathbf{k},\downarrow})$. The spin-orbit splitting leads to two bands with dispersion relations
\begin{equation}
E_{\mathbf{k},n}=N_{0,\mathbf{k}}+(-1)^n |\mathbf{N}_\mathbf{k}|,
\end{equation}
and the Bloch wavefunctions for these bands are respectively given by
\begin{equation}
\phi_+=\bigg(\begin{array}{c}
\cos \frac{\theta_\mathbf{k}}{2} \; e^{i \varphi_\mathbf{k}}  \\
\sin \frac{\theta_\mathbf{k}}{2}  \end{array} \bigg ), \: \: \: \: \: \phi_-=\bigg(\begin{array}{c}
-\sin \frac{\theta_\mathbf{k}}{2} \; e^{i \varphi_\mathbf{k}}  \\
\cos \frac{\theta_\mathbf{k}}{2}  \end{array} \bigg ),
\end{equation}
where $\cos \theta_{\mathbf{k}}=N_{3,\mathbf{k}}/|\mathbf{N}_\mathbf{k}|$, and $\tan \varphi_{\mathbf{k}}= N_{2,\mathbf{k}}/N_{1,\mathbf{k}}$. From these wavefunctions we immediately obtain the Berry curvatures by using the formula
\begin{eqnarray}\label{Berry}
\Omega_{\mathbf{k},n,a}=(-1)^n\epsilon_{abc}\frac{\mathbf{N}_{\mathbf{k}} \cdot \bigg[ \frac{\partial \mathbf{N}_{\mathbf{k}}}{\partial k_b}\times \frac{\partial \mathbf{N}_{\mathbf{k}}}{\partial k_c}\bigg]}{4|\mathbf{N}_{\mathbf{k}}|^3},
\end{eqnarray}
which shows that $\Omega_{\mathbf{k},n,a}$ is entirely determined by the spin-orbit coupling terms. The spin-orbit splitting $|\mathbf{N}_\mathbf{k}|$ vanishes at the following eight high symmetry points $\Gamma : (0,0,0)$; $M : (\pi, \pi, 0), \; (\pi, 0, \pi), \; (0, \pi, \pi)$; $X: (\pi, 0,0), \; (0, \pi, 0), \; (0,0,\pi)$; and $R: (\pi, \pi, \pi)$ of the cubic Brillouin zone, giving rise to the monopoles (right handed Weyl fermions) and the antimonopoles (left handed Weyl fermions) of $\Omega_{\mathbf{k},n,a}$. The explicit forms of the Berry curvature are provided in the Supplementary Material \cite{Supp}.

{\em Continuum limit :} Four flavors of right handed Weyl fermions are located at the $\Gamma$ and the $M$ points and their left handed counterparts are found at the $X$ and the $R$ points. When the spin independent hopping strengths are much smaller compared to $t_{SO}$, the following linearized Hamiltonian
\begin{eqnarray}
H &\approx& \sum^4_{\alpha=1}\Psi^{\dagger}_{\alpha, \mathbf{k}}\bigg[\epsilon_{\alpha,+}\sigma_0 \otimes \tau_0 +\epsilon_{\alpha,-}\sigma_0 \otimes \tau_3 + \nonumber \\&& v \mathbf{k} \cdot \boldsymbol \sigma \otimes \tau_3 \bigg]\Psi_{\alpha,\mathbf{k}},
\end{eqnarray}
produces an adequate description of the low energy physics, where $v=t_{SO}$. The $2 \times 2$ identity matrix $\tau_0$ and Pauli matrices $\boldsymbol \tau$ work on the chiral (left/right) indices. The four component spinors are defined as
\begin{equation}
\Psi^\dagger_{\alpha,\mathbf{k}}=(\psi^\dagger_{\mathbf{k}+\mathbf{Q}_{\alpha-1}},\psi^\dagger_{\mathbf{k}+\mathbf{Q}_{\alpha-1}+\mathbf{Q}^\ast}\sigma_{\alpha-1}),
\end{equation} where $\mathbf{Q}_0=(0,0,0)$, $\mathbf{Q}_1=(0, \pi, \pi)$, $\mathbf{Q}_2=(\pi,0,\pi)$, $\mathbf{Q}_3=(\pi, \pi, 0)$, and $\mathbf{Q}^\ast=(\pi,\pi,\pi)$. The energy shifting parameters are given by $\epsilon_{1,+}=12 t_2+6t_4$, $\epsilon_{1,-}=6t_1+8t_3$, $\epsilon_{2,+}=\epsilon_{3,+}=\epsilon_{4,+}=-4t_2+6t_4$, $\epsilon_{2,-}=\epsilon_{3,-}=\epsilon_{4,-}=-2t_1+8t_3$. More details about the linearization of the Hamiltonian can be found in the Supplementary Material\cite{Supp}.

The momentum separations for the Weyl points for the four flavors are actually given by $\delta \mathbf{K}_1=(-\pi, -\pi, -\pi)$, $\delta \mathbf{K}_2=(\pi, \pi, -\pi)$, $\delta \mathbf{K}_3=(\pi, -\pi, \pi)$ and $\delta \mathbf{K}_4=(-\pi, \pi, \pi)$. Notice, that $\sum_{\alpha} \delta \mathbf{K}_\alpha=0$, and there is no anomalous Hall effect \cite{Ran}. This is tied to the vanishing of the net Berry's flux through any plane, as a consequence of the time reversal symmetry. The sum of the energy shifts between the left and the right Weyl points, for all four flavors is $\sum_{\alpha}\epsilon_{\alpha,-}=64 t_3$. Therefore, a CME can arise when $t_3 \neq 0$. Otherwise, there will be a cancelation of the CME among the four flavors.

The finite AHE of a pair of Weyl fermions, as described in the introduction can be recovered, if we add a time reversal symmetry breaking perturbation
\begin{equation}
\Delta H= m(2-\cos k_1-\cos k_2)\; \sigma_3,
\end{equation}
to $H$ in Eq.~(\ref{tightbinding}), where $m>t_{SO}/2$. Under this perturbation only one pair of Weyl points at $(0,0,0)$ and $(0,0,\pi)$ survive with $\Delta \mathbf{K}=(0,0,-\pi)$, and provides a net Berry flux through the $xy$ plane. For such a model the CME exists, when either of $t_1$ and $t_3$ is nonzero.

It is important to note that the spin independent nearest neighbor hopping $t_1$ generally sets the largest energy scale for real materials, and the linearized theory is inadequate for their description. We must retain the higher gradient terms arising from $N_{0, \mathbf{k}}$, which are responsible for producing spin-orbit split Fermi pockets from both $\pm$ bands around these high symmetry points \cite{Takimoto}. However, the linearized theory provides us the understanding of the Berry phase effects in a simple manner, and also sets the upper bound for the anomalous conductivities.
\begin{figure*}[htb]
\centering
\subfigure[]{
\includegraphics[width=8.6cm,height=6cm]{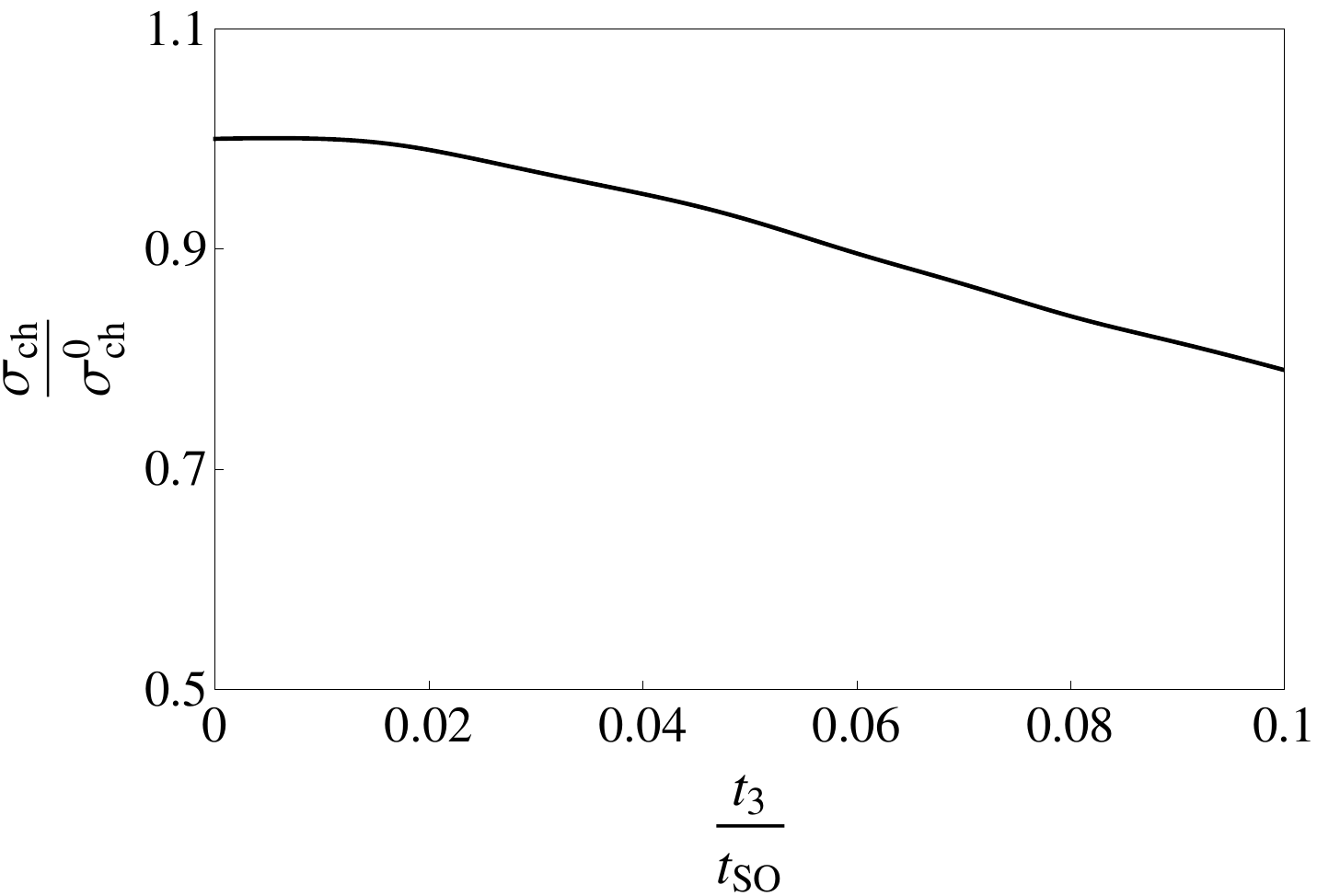}
\label{fig:subfig1a}
}
\subfigure[]{
\includegraphics[width=8.6cm,height=6cm]{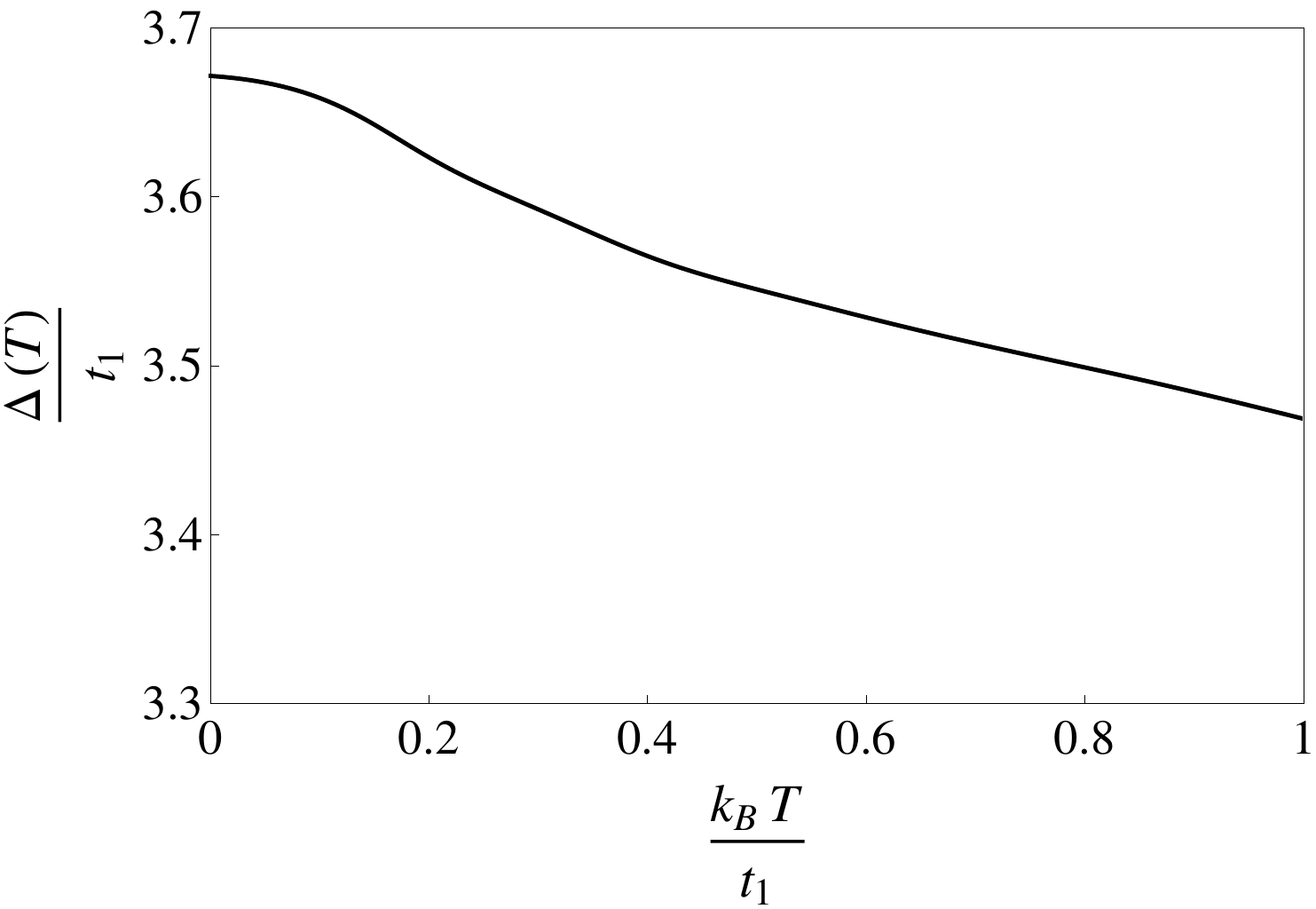}
\label{fig:subfig1b}
}
\label{fig:1}
\caption[]{(a)Normalized chiral magnetic conductivity $\sigma_{ch}/\sigma^0_{ch}$, as a function of $t_3/t_{SO}$, when $t_1=t_2=t_4=\mu=0$, where $\sigma^0_{ch}= e^2 (64 t_3)/h^2$ is the chiral magnetic conductivity, obtained from a linearized continuum theory. (b) Dimensionless energy parameter $\Delta(T)/t_1$ as a function of $k_B T/t_1$, when $t_2/t_1=-0.03$, $t_3/t_1=-0.88$, $t_4/t_1=-0.03$, and $\mu/t_1=0.02$, and the chiral magnetic conductivity is given by $\sigma_{ch}(T)= e^2 \Delta(T) /h^2$.
}
\end{figure*}

{\em Chiral magnetic conductivity: } For computing the chiral magnetic conductivity of a cubic noncentrosymmetric metal, we adopt the formalism of linear response theory. The information regarding the electrodynamic response is encoded in the current-current correlation function or the polarization tensor
\begin{eqnarray}
\Pi_{ab}(iq_0,\mathbf{q})&=&k_BT\; \sum_{k_0,\mathbf{k}} \; \mathrm{Tr}\bigg[j_a(\mathbf{k}) G\left(ik_0+\frac{iq_0}{2},\mathbf{k}+\frac{\mathbf{q}}{2}\right)  \nonumber \\ && \times j_b(\mathbf{k}) G\left(ik_0-\frac{iq_0}{2},\mathbf{k}-\frac{\mathbf{q}}{2}\right)\bigg],\label{polarization}
\end{eqnarray}
where \begin{equation}
G(ik_0, \mathbf{k})=\frac{(ik_0+\mu-N_{0,\mathbf{k}}) \tau_0+ \mathbf{N}_{\mathbf{k}} \cdot \boldsymbol \tau}{(ik_0+\mu-N_{0,\mathbf{k}})^2-\mathbf{N}^2_{\mathbf{k}}}.
\end{equation}
is the fermion propagator, and the paramagnetic current vertices are defined as $j_a=\frac{\partial H}{\partial k_a}$. We will only be interested in evaluating the antisymmetric parts of $\Pi_{ab}(iq_0,\mathbf{q})$. For that reason we only need to consider $\Pi^{ant}_{ab}(iq_0,\mathbf{q})=(\Pi_{ab}(iq_0,\mathbf{q})-\Pi_{ba}(iq_0,\mathbf{q}))/2$. The details of the calculation can be found in the Supplementary Material \cite{Supp}.

For the dc Hall conductivity ($\mathbf{q} \to 0$ before $q_0 \to 0$ as it is a response to a spatially uniform electric field), we have found
\begin{eqnarray}\label{AHE}
\sigma_{ab}=\frac{e^2}{\hbar} \; \epsilon_{abc} \; \sum_n \int \frac{d^3k}{(2\pi)^3} \Omega_{\mathbf{k},n,c} \; f(E_{\mathbf{k},n}),
\end{eqnarray}
where $f(E)=[\exp \{\beta (E-\mu)\}+1]^{-1}]$ is the Fermi distribution function, and $\beta=1/{k_BT}$. This is a well known formula for $\sigma_{ab}$ in terms of the Berry curvature \cite{Haldane}. For our model with time reversal symmetry, $\sigma_{ab}$ naturally vanishes. However, this model can support a $\mathbf{q}$-dependent Hall conductivity, which is important for finding the gyrotropic conductivity (a response to $\nabla \times \mathbf{E}=-\partial_t \mathbf{B}$) \cite{Mineev,Moore,Raghu}. However, we do not pursue this here.

For the chiral magnetic conductivity $j_{ch,a}/B_a$, which is a response to a time independent external magnetic field, it is important to consider the $\mathbf{q}$-linear terms of $\Pi^{ant}_{ab}(iq_0,\mathbf{q})$, and take the real frequency $q_0 \ll |\mathbf{q}|$ (opposite to what is done for the electric field). We find that the dc chiral magnetic conductivity is given by
\begin{eqnarray}\label{chmag1}
&&\sigma_{ch}=\frac{e^2}{\hbar^2} \; \int \frac{d^3k}{(2\pi)^3} \; \sum_n \bigg[\mathbf{v}_{\mathbf{k},n} \cdot \mathbf{\Omega}_{\mathbf{k},n} \; f(E_{\mathbf{k},n}) \nonumber \\ &&-(-1)^n |\mathbf{N}_{\mathbf{k}}| \; \mathbf{v}_{\mathbf{k},n} \cdot \mathbf{\Omega}_{\mathbf{k},n} \; f^\prime(E_{\mathbf{k},n})\bigg],
\end{eqnarray}
where $\mathbf{v}_{\mathbf{k},n}=\nabla_{\mathbf{k}}E_{\mathbf{k},n}$. Within a semiclassical derivation, only the first part involving the Fermi functions is found, and this part vanishes after the sum over the band index $n$ and the integrals over momentum are carried out\cite{Basar,Niu,Yamamoto,Stephanov}. The nonzero $\sigma_{ch}$ entirely comes from the second term, involving the derivative of the Fermi functions (which at $T=0$ reduce to delta functions, and define the underlying Fermi surfaces). When the above formula is used for the left and the right handed fermions of the continuum model of Eq.~(\ref{eq:1}), the result of Eq.~(\ref{eq:3}) are correctly reproduced. If we rather consider a massive Dirac fermion in the presence of an axial chemical potential (by setting $\delta \mathbf{K}=0$ in Eq.~(\ref{eq:1}), and adding a spin independent hybridization between the left and the right fermions), the spectrum remains fully gapped, and the CME vanishes due to the absence of Fermi surfaces \cite{Goswami1}. In contrast, a different formula
\begin{eqnarray}\label{chmag2}
&&\tilde{\sigma}_{ch}=\frac{e^2}{\hbar^2} \; \int \frac{d^3k}{(2\pi)^3} \; \sum_n \bigg[\frac{\mathbf{v}_{\mathbf{k},+}+\mathbf{v}_{\mathbf{k},-}}{2} \cdot \mathbf{\Omega}_{\mathbf{k},n} \; f(E_{\mathbf{k},n}) \nonumber \\ && -\frac{(-1)^n}{3} |\mathbf{N}_{\mathbf{k}}| \; \mathbf{v}_{\mathbf{k},n} \cdot \mathbf{\Omega}_{\mathbf{k},n} \; f^\prime(E_{\mathbf{k},n})\bigg],
\end{eqnarray}
is obtained in the opposite limit $|\mathbf{q}| \ll q_0$. Even though this formula predicts a nonzero answer, it does not reproduce the result of Eq.~(\ref{eq:3}) in the continuum limit. Since, we are trying to calculate the current in response to a static magnetic field, $q_0 \ll |\mathbf{q}|$ is the appropriate limit and Eq.~(\ref{chmag1}) is the correct formula for the chiral magnetic conductivity. Remarkably Eq.~(\ref{chmag1}) can be applied to a generic band structure, and clarifies the important role of the Berry curvature.

For obtaining $\sigma_{ch}$, we have performed the integrals over the momentum components in Eq.~(\ref{chmag1})numerically. We first choose a simple limit $t_1=t_2=t_4=\mu=0$ (this is an unphysical choice for real materials), and study the dependence of $\sigma_{ch}$ at zero temperature, on $t_3/t_{SO}$. In this case the continuum limit suggests $\sigma^0_{ch}= e^2 (64 t_3)/h^2$. We find departure from the continuum limit answer, when $t_3/t_{SO}> 0.02$. But, it is important to note that the $\sigma_{ch}$ is finite, even when the linearized continuum theory is invalid. This is demonstrated in Fig.~\ref{fig:subfig1a}. In addition, we have also studied the temperature dependence of $\sigma_{ch}$, by employing Eq.~(\ref{chmag1}).  We can capture the temperature dependence in terms of an energy parameter $\Delta(T)$, such that $\sigma_{ch}(T)=e^2 \Delta(T)/h^2$. In Fig.~\ref{fig:subfig1b}, we show $\Delta(T)$ as function of $T$, for the band parameters $t_2/t_1=-0.03$, $t_3/t_1=-0.88$, $t_4/t_1=-0.03$, $\mu/t_1=0.02$.This parametrization has been used in Ref.~\onlinecite{Takimoto}, for fitting the fermiology of Li$_2$Pt$_3$B, obtained from the ab initio calculations \cite{Chandra,Pickett}. Even for $k_B T/t_1$ comparable to unity, we find a substantially large $\sigma_{ch}$, and $\sigma_{ch} \sim e^2 (0.09 t_1)/h^2$.

In conclusion, our calculations demonstrate the existence of the CME in a lattice model. We have derived a formula for the chiral magnetic conductivity in Eq.~(\ref{chmag1}), which is applicable for a generic band structure of the noncentrosymmetric materials. In addition, we have also identified the cubic noncentrosymmetric metals as hosts of Weyl points, occurring at different reference energies. We have found a substantial chiral magnetic conductivity even at the room temperature. If we choose $t_1 \sim 1 eV$, $T=300 K$, we roughly obtain $j_{ch} \approx 3.3$ A/mm$^2$ for $B= 1$ Gauss, which is measurable with conventional experimental set up. Hence, our work identifies the possibility of realizing the exotic CME in real materials, which can have important applications for developing magnetic field sensors.

{\em Acknowledgment:} Work supported by the NSF Cooperative Agreement No.DMR-
0654118, the State of Florida, and the U. S. Department
of Energy (P.G.) and NSF (PHY-1104527) and AFOSR (FA9550-13-1-0045) (S.T.)

\newpage

\onecolumngrid

\section*{Supplementary Material for EPAPS}

\subsection{I.Explicit forms of the Berry curvature}
After applying the formula in Eq.~(\ref{Berry}) for the tight-binding Hamiltonian of Eq.~(\ref{tightbinding}), we obtain
\begin{eqnarray}
\Omega_{\mathbf{k},n,1}=\frac{(-1)^n} {2|\mathbf{N}_\mathbf{k}|^3}N_{1,\mathbf{k}} \frac{\partial N_{2,\mathbf{k}}}{ \partial k_2}\frac{\partial N_{3,\mathbf{k}}}{ \partial k_3}=(-1)^n \frac{\sin k_1 \cos k_2 \cos k_3}{2\left[\sin^2 k_1 +\sin^2 k_2 +\sin^2 k_3\right]^{\frac{3}{2}}}, \\
\Omega_{\mathbf{k},n,2}=\frac{(-1)^n} {2|\mathbf{N}_\mathbf{k}|^3}N_{2,\mathbf{k}} \frac{\partial N_{3,\mathbf{k}}}{ \partial k_3}\frac{\partial N_{1,\mathbf{k}}}{ \partial k_1}=(-1)^n \frac{\sin k_2 \cos k_1 \cos k_3}{2\left[\sin^2 k_1 +\sin^2 k_2 +\sin^2 k_3\right]^{\frac{3}{2}}},\\
\Omega_{\mathbf{k},n,3}=\frac{(-1)^n} {2|\mathbf{N}_\mathbf{k}|^3}N_{3,\mathbf{k}} \frac{\partial N_{1,\mathbf{k}}}{ \partial k_1}\frac{\partial N_{2,\mathbf{k}}}{ \partial k_2}=(-1)^n \frac{\sin k_3 \cos k_1 \cos k_2}{2\left[\sin^2 k_1 +\sin^2 k_2 +\sin^2 k_3\right]^{\frac{3}{2}}},
\end{eqnarray}
Around $\mathbf{Q}_0=(0,0,0)$ and $\mathbf{Q}^\ast=(\pi, \pi, \pi)$, the Berry curvatures respectively behave as
\begin{eqnarray}
\Omega_{\mathbf{Q}_0+\mathbf{k},n,a}=(-1)^n \frac{k_a}{2|\mathbf{k}|^3}, \; \; \; \Omega_{\mathbf{Q}^\ast+\mathbf{k},n,a}=(-1)^{n+1} \frac{k_a}{2|\mathbf{k}|^3},
\end{eqnarray}
which explicitly show that the right and the left Weyl points respectively act as the monopole and the antimonopole of the Berry curvature.

\subsection{II. Continuum limit  of the tight-binding model}
If we expand the tight-binding Hamiltonian of Eq.~(\ref{tightbinding}) around the low energy points at $\mathbf{Q}_0=(0,0,0)$ and $\mathbf{Q}^\ast=(\pi,\pi,\pi)$ up to the quadratic order in $\mathbf{k}$, we obtain
\begin{eqnarray}
&&h_{0,0,0}= 2t_1\left(3-\frac{\mathbf{k}^2}{2}\right)+4t_2(3-\mathbf{k}^2)+8t_3\left(1-\frac{\mathbf{k}^2}{2}\right)+2t_4(3-2\mathbf{k}^2)+t_{SO}\mathbf{k} \cdot \boldsymbol \sigma, \label{continuum1} \\
&&h_{\pi,\pi,\pi}=-2t_1\left(3-\frac{\mathbf{k}^2}{2}\right)+4t_2(3-\mathbf{k}^2)-8t_3\left(1-\frac{\mathbf{k}^2}{2}\right)+2t_4(3-2\mathbf{k}^2)-t_{SO}\mathbf{k} \cdot \boldsymbol \sigma. \label{continuum2}
\end{eqnarray}
Similarly around $\mathbf{Q}_3=(\pi, \pi,0)$ and $\mathbf{Q}_3+\mathbf{Q}^\ast \equiv (0,0,\pi)$ we find
\begin{eqnarray}
&&h_{\pi,\pi,0}= -2t_1\left(1-\frac{\mathbf{k}^2-2k^2_3}{2}\right)-4t_2(1-k^2_3)+8t_3\left(1-\frac{\mathbf{k}^2}{2}\right)+2t_4(3-2\mathbf{k}^2)-t_{SO}\mathbf{k} \left(k_1 \sigma_1+k_2 \sigma_2 -k_3 \sigma_3\right), \label{continuum3}\\
&&h_{0,0,\pi}=2t_1\left(1-\frac{\mathbf{k}^2-2k^2_3}{2}\right)-4t_2(1-k^2_3)-8t_3\left(1-\frac{\mathbf{k}^2}{2}\right)+2t_4(3-2\mathbf{k}^2)+t_{SO}\mathbf{k} \left(k_1 \sigma_1+k_2 \sigma_2 -k_3 \sigma_3\right).\label{continuum4}
\end{eqnarray}
After a unitary transformation $\psi \to \sigma_3 \psi$, the spin orbit contributions around these two points become $\pm t_{SO}\mathbf{k} \cdot \boldsymbol \sigma$. Around $\mathbf{Q}_2=(\pi, 0,\pi)$ and $\mathbf{Q}_2+\mathbf{Q}^\ast \equiv (0,\pi,0)$, the low energy Hamiltonians are respectively obtained by changing $k_3$ to $k_2$ in Eq.~(\ref{continuum3}) and Eq.~(\ref{continuum4}), and the unitary transformation $\psi \to \sigma_2 \psi$ change the spin-orbit parts to $\pm t_{SO}\mathbf{k} \cdot \boldsymbol \sigma$. For the low energy points at $\mathbf{Q}_1=(0,\pi,\pi)$ and $\mathbf{Q}_1+\mathbf{Q}^\ast \equiv (\pi,0,0)$, we replace $k_3$ by $k_1$ in Eq.~(\ref{continuum3}) and Eq.~(\ref{continuum4}), and perform the unitary transformation $\psi \to \sigma_1 \psi$.

In the absence of spin independent hopping terms, there is a nesting symmetry $\mathbf{N}_{\mathbf{k}+\mathbf{Q}^\ast}=-\mathbf{N}_{\mathbf{k}}$, where $\mathbf{Q}^\ast=(\pi, \pi, \pi)$. This generates the classical chiral symmetry (with respect to $\gamma_5$) between the right and the left handed fermions. In addition, if we let $\mathbf{k} \to \mathbf{k}+\mathbf{Q}_j$, with $j=1,2,3$, and $\mathbf{Q}_1=(0, \pi, \pi)$, $\mathbf{Q}_2=(\pi,0,\pi)$, $\mathbf{Q}_3=(\pi, \pi, 0)$, only two components of $\mathbf{N}_{\mathbf{k}}$ change signs. The sign changes can be compensated by the subsequent redefinitions of the spinors according to $\psi_{\mathbf{k}+\mathbf{Q}_j} \to \sigma_{j} \psi_{\mathbf{k}+\mathbf{Q}_j}$. This is how the non-Abelian flavor symmetry emerges. The spin independent hopping terms destroy these nesting properties, and introduce flavor dependent energy shifts between the left and the right handed Weyl points. Due to the absence of any translational symmetry breaking order with wave-vector $\mathbf{Q}^\ast$, the $U(1)$ chiral symmetry with respect to $\gamma_5$ is still preserved. Similarly, the flavor symmetry is not completely eliminated either, as we are not considering any translational symmetry breaking orders with the wave vectors $\mathbf{Q}_1$, $\mathbf{Q}_2$ and $\mathbf{Q}_3$, which lead to the mixing between different flavors. Due to the flavor dependent energy separations, the flavor symmetry is just lowered to a smaller subgroup.\\

\subsection{III. Details of the linear response calculation}
There are three types of Berry phase controlled electrodynamic response, which can arise from the antisymmetric part of the polarization tensor in a noncentrosymmetric system: (i) the AHE, with respect to an external, spatially uniform electric field, (ii) the CME with respect to an external, time independent magnetic field, and (iii) the gyrotropic response with respect to $\nabla \times \mathbf{E}= -\frac{\partial \mathbf{B}}{\partial t}$. Here, we will confine ourselves to the discussion of only the AHE and the CME. The corresponding response functions or the conductivities are respectively defined as
\begin{eqnarray}
\sigma_{ab}=\lim_{q_0 \to 0}\frac{1}{iq_0} \lim_{\mathbf{q} \to 0} \Pi^{ant}_{ab,R}(q_0,\mathbf{q}), \\
\sigma_{ch}= \lim_{\mathbf{q} \to 0} \epsilon_{abc}\frac{1}{2iq_c} \lim_{q_0 \to 0}\Pi^{ant}_{ab,R}(q_0,\mathbf{q}),
\end{eqnarray}
where the subscript $R$ signifies the retarded correlation function (after the analytic continuation of the Matsubara frequency to the real frequency $iq_0 \to q_0 + i\delta$). The reason for these definitions are of course tied to the definition of the electromagnetic field strengths $\mathbf{E}= -\frac{\partial \mathbf{A}}{\partial t}$ and $\mathbf{B}=\nabla \times \mathbf{A}$. For the gyrotropic response we need to divide by $q_0 q_c$.

For concreteness we here focus on $\Pi^{ant}_{23,R}(q_0,\mathbf{q})$, where $\mathbf{q}=(q_1,0,0)$. The denominator of the integrand for the polarization tensor in Eq.~(\ref{polarization})is given by
\begin{eqnarray}
\mathcal{D}(ik_0,iq_0,\mathbf{k}, \mathbf{q})=\left[\left \{ik_0+iq_0+\mu-N_{0,\mathbf{k}+\mathbf{q}/2}\right \}^2-\mathbf{N}^2_{\mathbf{k}+\mathbf{q}/2}\right]\left[\left \{ik_0+\mu-N_{0,\mathbf{k}-\mathbf{q}/2}\right \}^2-\mathbf{N}^2_{\mathbf{k}-\mathbf{q}/2}\right].
\end{eqnarray}
The numerator has the following expression
\begin{eqnarray}
&&\mathcal{N}_{23}(ik_0,iq_0,\mathbf{k}, \mathbf{q})=-2i\bigg[\frac{\partial N_{2,\mathbf{k}}}{\partial k_2} \frac{\partial N_{3,\mathbf{k}}}{\partial k_3}\bigg(\left \{ik_0+\mu-N_{0,\mathbf{k}-\mathbf{q}/2}\right \}N_{1,\mathbf{k}+\mathbf{q}/2}-\left \{ik_0+iq_0+\mu-N_{0,\mathbf{k}+\mathbf{q}/2}\right \}N_{1,\mathbf{k}-\mathbf{q}/2}\bigg) \nonumber \\ &&
 + \frac{\partial N_{0,\mathbf{k}}}{\partial k_2} \frac{\partial N_{3,\mathbf{k}}}{\partial k_3}\bigg(N_{1,\mathbf{k}+\mathbf{q}/2}N_{2,\mathbf{k}-\mathbf{q}/2}- N_{1,\mathbf{k}-\mathbf{q}/2}N_{2,\mathbf{k}+\mathbf{q}/2}\bigg)+ \frac{\partial N_{0,\mathbf{k}}}{\partial k_3} \frac{\partial N_{2,\mathbf{k}}}{\partial k_2}\bigg(N_{1,\mathbf{k}+\mathbf{q}/2}N_{3,\mathbf{k}-\mathbf{q}/2}- N_{1,\mathbf{k}-\mathbf{q}/2}N_{3,\mathbf{k}+\mathbf{q}/2}\bigg)
\bigg]. \nonumber \\
\end{eqnarray}
We first perform a Taylor expansion of the numerator in powers of the external momentum $\mathbf{q}$, and retain only the $q$ independent and the $q$-linear terms to obtain
\begin{eqnarray}
&&\mathcal{N}_{23}(ik_0,iq_0,\mathbf{k}, \mathbf{q}) \approx -2i\bigg[-iq_0 \; N_{1,\mathbf{k}}\frac{\partial N_{2,\mathbf{k}}}{\partial k_2} \frac{\partial N_{3,\mathbf{k}}}{\partial k_3}+q_1 \bigg(\frac{\partial N_{0,\mathbf{k}}}{\partial k_2} \; N_{2, \mathbf{k}} \frac{\partial N_{3,\mathbf{k}}}{\partial k_3} \frac{\partial N_{3,\mathbf{k}}}{\partial k_3} +  \frac{\partial N_{0,\mathbf{k}}}{\partial k_3} \; N_{3, \mathbf{k}} \frac{\partial N_{1,\mathbf{k}}}{\partial k_1} \frac{\partial N_{2,\mathbf{k}}}{\partial k_2}\nonumber \\
&&+\frac{\partial N_{0,\mathbf{k}}}{\partial k_1}N_{1,\mathbf{k}}\frac{\partial N_{2,\mathbf{k}}}{\partial k_2} \frac{\partial N_{3,\mathbf{k}}}{\partial k_3}\bigg)+q_1\frac{\partial N_{1,\mathbf{k}}}{\partial k_1}\frac{\partial N_{2,\mathbf{k}}}{\partial k_2} \frac{\partial N_{3,\mathbf{k}}}{\partial k_3}\bigg(ik_0+\mu-N_{0,\mathbf{k}}+i\frac{q_0}{2}\bigg)\bigg].
\end{eqnarray}

For obtaining the anomalous Hall conductivity, we can set $\mathbf{q}=0$ both in the numerator and the denominator. The numerator of the polarization tensor now has the following form
\begin{equation}
\mathcal{N}_{ab}(ik_0, \mathbf{k},iq_0,\mathbf{q}=0)=2 i \; \epsilon_{abc} \sum_n (-1)^n \Omega_{\mathbf{k},n,c} |\mathbf{N}_{\mathbf{k}}|^3 \; (iq_0)
\end{equation}
After performing the Matsubara sum, we analytically continue to the real frequencies, and divide by $i q_0$. The real part of the polarization tensor leads to the following formula for the dynamic anomalous Hall conductivity
\begin{eqnarray}
\sigma_{ab}(q_0)=4\epsilon_{abc}\sum_n \int \frac{d^3k}{(2\pi)^3} \frac{\Omega_{\mathbf{k},n,c} \; \mathbf{N}^2_{\mathbf{k}} \; f(E_{n,\mathbf{k}})}{4\mathbf{N}^2_{\mathbf{k}}-q^2_0}.
\end{eqnarray}
After setting $q_0=0$, we arrive at Eq.~(\ref{AHE}).

For obtaining the chiral magnetic conductivity, it is also important to retain the momentum dependence of the denominator to the linear order. However, instead of linearizing the denominator, we first perform the Matsubara sum, and subsequently expand in powers of $q_1$ to arrive at
\begin{eqnarray}
&&\Pi^{ant}_{23}(iq_0,q_1)\approx \frac{i}{2} \bigg[-\tanh \frac{\beta \xi_{\mathbf{k},+}}{2}\bigg \{q_1\mathbf{v}_{\mathbf{k},+}\cdot \mathbf{\Omega}_{\mathbf{k},+}-iq_0\Omega_{\mathbf{k},+,1}-\nabla_{\mathbf{k}}|\mathbf{N}_{\mathbf{k}}|\cdot \mathbf{\Omega}_{\mathbf{k},+} \frac{iq_1q_0}{iq_0-q_1v_{\mathbf{k},+,1}}\bigg \}-\tanh \frac{\beta \xi_{\mathbf{k},-}}{2}\times\nonumber \\&& \bigg \{q_1\mathbf{v}_{\mathbf{k},-}\cdot \mathbf{\Omega}_{\mathbf{k},-}-iq_0\Omega_{\mathbf{k},-,1}+\nabla_{\mathbf{k}}|\mathbf{N}_{\mathbf{k}}|\cdot \mathbf{\Omega}_{\mathbf{k},-} \frac{iq_1q_0}{iq_0-q_1v_{\mathbf{k},-,1}}\bigg \} -\mathrm{sech}^2 \frac{\beta \xi_{\mathbf{k},+}}{2} \; \frac{\beta |\mathbf{N}_{\mathbf{k}}|}{2} q_1v_{\mathbf{k},+,1}\frac{q_1\mathbf{v}_{\mathbf{k},+}\cdot \mathbf{\Omega}_{\mathbf{k},+}-iq_0\Omega_{\mathbf{k},+,1}}{iq_0-q_1v_{\mathbf{k},+,1}}\nonumber \\ && + \; \mathrm{sech}^2 \frac{\beta \xi_{\mathbf{k},-}}{2} \; \frac{\beta |\mathbf{N}_{\mathbf{k}}|}{2} q_1v_{\mathbf{k},-,1}\frac{q_1\mathbf{v}_{\mathbf{k},-}\cdot \mathbf{\Omega}_{\mathbf{k},-}-iq_0\Omega_{\mathbf{k},-,1}}{iq_0-q_1v_{\mathbf{k},-,1}}\bigg]
\end{eqnarray}
where $\xi_{\mathbf{k},\pm}=E_{\mathbf{k},\pm}-\mu$, and $\mathbf{v}_{\mathbf{k},\pm}=\nabla_{\mathbf{k}}E_{\mathbf{k},\pm}$. After analytically continuing to real frequencies, we first consider $q_0 \ll |\mathbf{q}|$. After setting $q_0=0$, and dividing by $iq_1$, we obtain the formula in Eq.~(\ref{chmag1}). In the opposite limit, $|\mathbf{q}| \ll q_0$, after dividing by $iq_1$, the frequency independent part leads to
the formula in Eq.~(\ref{chmag2}).

\begin{thebibliography}{10}

\bibitem{Volovik1} G. E. Volovik, {\it Universe in a helium droplet} (Oxford University Press, 2003).

\bibitem{Vishwanath} X. Wan, A. Turner, A. Vishwanath, and S. Y. Savrasov, Phys. Rev. B \textbf{83}, 205101 (2011).

\bibitem{Xu} G. Xu, H. Weng, Z. Wang, X. Dai, and Z. Fang, Phys. Rev. Lett. \textbf{107}, 186806 (2011).

\bibitem{Burkov1} A. A. Burkov, and L. Balents, Phys. Rev. Lett. \textbf{107}, 127205 (2011).

\bibitem{Burkov2} A. A. Burkov, M. D. Hook, and L. Balents, Phys. Rev. B \textbf{84}, 235126 (2011).

\bibitem{Zyuzin1} A. A. Zyuzin, S. Wu, and A. A. Burkov, Phys. Rev. B \textbf{85}, 165110 (2012).

\bibitem{Zyuzin2} A. A. Zyuzin, A. A. Burkov, Phys. Rev. B \textbf{86}, 115133 (2012).

\bibitem{Cho} G. Y. Cho, arXiv:1110.1939 (2011).

\bibitem{Meng} T. Meng, and L. Balents, Phys. Rev. B \textbf{86}, 054504 (2012).

\bibitem{Gong} M. Gong, S. Tewari, C. W. Zhang, Phys. Rev. Lett. \textbf{107}, 195303 (2011).

\bibitem{Sau} J. D. Sau, S. Tewari, Phys. Rev. B \textbf{86}, 104509 (2012).

\bibitem{Das} T. Das, Phys. Rev. B \textbf{88}, 035444 (2013).

\bibitem{Goswami1} P. Goswami, and S. Chakravarty, Phys. Rev. Lett. \textbf{107}, 196803 (2011).

\bibitem{Hosur} P. Hosur, S. A. Parameswaran, and A. Vishwanath,  Phys. Rev. Lett. 108, 046602 (2012).

\bibitem{Zyuzin3} A. A. Zyuzin, and A. A. Burkov, Phys. Rev. B \textbf{86}, 115133 (2012).

\bibitem{Grushin} A. G. Grushin, Phys. Rev. D \textbf{86}, 045001 (2012).

\bibitem{Qi} C. X. Liu, P. Ye, X. L. Qi, Phys. Rev. B \textbf{87}, 235306 (2013).

\bibitem{Goswami2} P. Goswami, and S. Tewari, arXiv:1210.6352

\bibitem{Kharzeev} K. Fukushima, D. E. Kharzeev, and H. J. Warringa,  Phys. Rev. D \textbf{78}, 074033 (2008).

\bibitem{Landsteiner}K. Landsteiner, E. Megias, and F. Pena-Benitez, Phys. Rev. Lett. \textbf{107}, 021601 (2011).

\bibitem{Franz}M.M. Vazifeh, M. Franz, Phys. Rev. Lett. \textbf{111}, 027201 (2013).

\bibitem{Basar}G. Basar, D. E. Kharzeev, H-U. Yee, arXiv:1305.6338

\bibitem{Landsteiner2} K. Landsteiner, arXiv:1306.4932

\bibitem{Burkov4} Y. Chen, Si Wu, A.A. Burkov, Phys. Rev. B \textbf{88}, 125105 (2013).

\bibitem{Fujikawa} K. Fujikawa, and H. Suzuki, {\it Path integrals and quantum anomalies} (Oxford University Press 2004).

\bibitem{Goswami3} P. Goswami and B. Roy, arXiv:1211.4023

\bibitem{Ninomiya}H. B. Nielsen and M. Ninomiya, Phys. Lett. B \textbf{130},
389 (1983).

\bibitem{Aji}V. Aji, Phys. Rev. B \textbf{85}, 241101 (2012).

\bibitem{Takimoto} S. P. Mukherjee, and T. Takimoto, Phys. Rev. B, \textbf{86}, 134526 (2012).

\bibitem{Supp} For details see Supplementary Material at EPAPS Document No. X-XXX-XXXXXX-XXX-XXXXXX.

\bibitem{Ran} K. Y. Yang, Y. M. Lu, and Y. Ran, Phys. Rev. B \textbf{84}, 075129 (2011).

\bibitem{Haldane} F. D. Haldane, Phys. Rev. Lett. \textbf{93}, 206602 (2004).

\bibitem{Mineev} V. P. Mineev and Y. Yoshioka, Phys. Rev. B \textbf{81}, 094525 (2010).

\bibitem{Moore}J. Orenstein, and J. E. Moore, Phys. Rev. B \textbf{87}, 165110 (2013).

\bibitem{Raghu} Pavan Hosur, A. Kapitulnik, S. A. Kivelson, J. Orenstein, and S. Raghu,
Phys. Rev. B \textbf{87}, 115116 (2013).

\bibitem{Niu}J. Zhou, H. Jiang, Q. Niu, and J. Shi, Chinese Phys. Lett. \textbf{30}, 027101 (2013).

\bibitem{Yamamoto}D. T. Son and N. Yamamoto, Phys. Rev. Lett. \textbf{109}, 181602 (2012).

\bibitem{Stephanov} M. A. Stephanov and Y. Yin, Phys. Rev. Lett. \textbf{109}, 162001 (2012).

\bibitem{Chandra} S. Chandra, S. Mathi Jaya, M. C. Valsakunmar, Physica C \textbf{432}, 116 (2005).

\bibitem{Pickett} K. W. Lee, and W. E. Pickett, Phys. Rev. B, \textbf{72}, 174505 (2005).

\end{thebibliography}
\end{document}